\documentclass[aps,prd,twocolumn,floatfix,superscriptaddress]{revtex4}
\usepackage{graphicx,amssymb,url}

\newcommand{\be}{\begin{equation}}
\newcommand{\ee}{\end{equation}}
\newcommand{\ba}{\begin{eqnarray}}
\newcommand{\ea}{\end{eqnarray}}

\def\lsim{\raise0.3ex\hbox{$\;<$\kern-0.75em\raise-1.1ex\hbox{$\sim\;$}}}
\def\gsim{\raise0.3ex\hbox{$\;>$\kern-0.75em\raise-1.1ex\hbox{$\sim\;$}}}
\def\eps{\varepsilon}
\def\theta{\vartheta}
\def\ap{\approx}

\begin{document}

\title{Restricting UHECRs and cosmogenic neutrinos with Fermi-LAT}

\author{V.~Berezinsky}
\affiliation{INFN, Laboratori Nazionali del Gran Sasso, I--67010
 Assergi (AQ), Italy}

\author{A.~Gazizov}
\affiliation{INFN, Laboratori Nazionali del Gran Sasso, I--67010
 Assergi (AQ), Italy}
\affiliation{Institutt for fysikk, NTNU, Trondheim, Norway}

\author{M.~Kachelrie\ss}
\affiliation{Institutt for fysikk, NTNU, Trondheim, Norway}

\author{S.~Ostapchenko}
\affiliation{Institutt for fysikk, NTNU, Trondheim, Norway}
\affiliation{D.~V.~Skobeltsyn Institute of Nuclear Physics,
Moscow State University, Russia}

\date{March 6, 2010}

\begin{abstract}
Ultrahigh energy cosmic ray (UHECR) protons interacting with the cosmic
microwave background (CMB)  produce UHE electrons and gamma-rays
that in turn initiate electromagnetic cascades on CMB and infrared photons.
As a result, a background of  diffuse isotropic  gamma
radiation is accumulated in the energy range $E\lsim 100$\,GeV.
The Fermi-LAT collaboration has recently reported a measurement
of the extragalactic diffuse background finding it
less intense and softer than previously measured by EGRET.
We show that this new result constrains UHECR models  and the flux of
cosmogenic neutrinos. In particular,
it excludes models with cosmogenic neutrino fluxes detectable by existing
neutrino experiments, while next-generation detectors as e.g.\ 
JEM-EUSO can observe neutrinos only for extreme parameters. 
\end{abstract}


\maketitle

\section{Introduction}%
The origin of ultrahigh energy cosmic rays (UHECRs)
is not yet established despite more than 50~years of research.
Natural candidates as UHECR primaries are extragalactic protons from
astrophysical sources. In this case, interactions of UHE protons with
the cosmic microwave background (CMB) leave their imprint on the UHECR
energy spectrum in the form of the Greisen-Zatsepin-Kuzmin (GZK)
cutoff and a pair-production dip~\cite{GZK}.

The GZK cutoff is a steepening of the proton spectrum at the energy
$E_{\rm GZK} \approx (4 - 5)\times 10^{19}$~eV, caused by photo-pion
production on the CMB. Such a steepening has been observed by the
HiRes~\cite{Abbasi:2007sv} and the Auger collaboration~\cite{Abraham:2008ru},
but its real cause is still unclear.

An immediate consequence of the dominance of extragalactic
protons in the CR flux and their interaction with CMB photons is the
existence of ultrahigh energy (``cosmogenic'') neutrinos produced
by charged pion decays, as suggested first in Ref.~\cite{BZ}.

Another signature for extragalactic protons is a pair-production 
dip \cite{BG88,BGG-dip} in the CR flux around $5\times 10^{18}$\,eV, 
which is clearly seen in the experimental data. Photons and positrons 
from pion decay and $p + \gamma_{\rm CMB} \to p + e^+ + e^-$ pair-production
initiate electromagnetic cascades on  photons from the CMB and the 
extragalactic background light (EBL), dumping all
the energy injected into cascade particles below the pair-production
threshold at $\sim 100$\,GeV.

Clearly, the production of neutrinos by UHE protons
is thus intimately tied to the one of photons and electrons, and both
depend in turn on the flux of primary cosmic rays.
While UHE photons and electrons start electromagnetic cascades by
scattering on photons from the EBL,
neutrinos reach us suffering no collisions. Therefore, the
measurement of the diffuse extragalactic gamma-ray background (EGRB)
can be used to impose a strict upper limit on the possible diffuse high energy
neutrino flux, as suggested first in Ref.~\cite{cascade}.

We derive in this work an upper limit on the flux of cosmogenic neutrinos 
assuming that the primary UHECR particles are protons. In case that all 
primaries or part of them are nuclei, the cosmogenic neutrino flux is
lower than for a pure proton composition~\cite{nuclei}. Thus our 
assumption of a pure proton composition is justified, since we aim at
deriving an {\em upper limit\/} on the cosmogenic neutrino flux.  
Note that the HiRes data~\cite{Abbasi:2007sv}  agree with a pure proton 
composition at $E \gsim 1\times 10^{18}$~eV, while the mass composition
deduced from the Auger data indicates the presence of heavier nuclei
in the primary UHECR flux~\cite{Abraham:2008ru}. In the latter case,
the maximally allowed cosmogenic neutrino flux would be below
the upper limit derived for a pure proton primary flux in this paper.

In the present work, we use a recently reported measurement~\cite{fermi}
of the EGRB by Fermi-LAT to constrain UHECR models.
We show that the observed fast decrease of the EGRB with
energy, $J(E)\propto E^{-2.41}$, already constrains such models. 
In particular, versions of the dip model with strong
redshift evolution contradict the Fermi data, while this model
without or with weak redshift evolution remains viable. Moreover,
the Fermi data allows us to derive a strong upper limit on the diffuse
UHE neutrino flux. As a result, we conclude that the detection of
cosmogenic neutrinos requires to increase the sensitivity of UHE neutrino
experiments compared to current levels.  
As it is demonstrated below,
the maximal energy density of cascade radiation
$\omega_{\rm cas}^{\max} \approx 5.8\times 10^{-7}$~eV/cm$^3$
allowed by the Fermi-LAT data can be used to select
viable UHECR models without explicitly  calculating
electromagnetic cascade processes. 

\section{Analytical calculations}%
The two basic processes driving an electromagnetic cascade  are pair
production (PP)   $\gamma\gamma_{\rm b}\to e^+e^-$ and inverse Compton (IC)
scattering $e^\pm\gamma_{\rm b}\to e^\pm\gamma$ on background photons
$\gamma_{\rm b}$. The cascade develops very fast
with a minimal interaction length $l_{\rm int}(E)\sim 10$\,kpc
until it reaches the pair creation threshold. From that point on, electrons
emit photons in the Thomson regime while photons stop interacting.
Their spectrum can be estimated analytically \cite{cascade,em-cascade}
in terms of the production rate of
the cascade photons $Q_{\gamma}^{\rm cas}(E)$ per unit volume as
\begin{equation}
\label{Qgamma}
Q_\gamma^{\rm cas}(E) = \left\{
\begin{array}{lll}
K (E/\varepsilon_X)^{-3/2}  \quad &\mbox{for }  &E \leq \varepsilon_X\,, \\
K (E/\varepsilon_X)^{-2}  \quad &\mbox{for }  &\varepsilon_X \leq E\leq
\varepsilon_a\,,
\end{array}
\right.
\end{equation}
with a steepening at $E>\varepsilon_a$. Here, $\varepsilon_a$ is
the minimal absorption energy of a cascade photon scattering on the EBL, and
$\varepsilon_X$ is the energy of a photon emitted by an electron/positron
($e_X + \gamma \rightarrow e' + \gamma_X$), which is in turn produced by a
photon $\gamma_a$ (via $\gamma_a + \gamma_{\rm EBL} \rightarrow e_X^+ + e_X^-$)
with the minimal absorption energy $\varepsilon_a$. The energy spectrum
(\ref{Qgamma}) of the cascade radiation typically extends
up to $\sim 100$\,GeV. The constant $K$ in Eq.~(\ref{Qgamma}) defines the
normalization of the production rate via $K = Q_\gamma^{\rm cas}(\varepsilon_X)$.
The two energies $\varepsilon_a$ and $\varepsilon_X$ are related to each
other as $\varepsilon_X = 1/3\left( \varepsilon_a/m_e \right)^2
\varepsilon_{\rm cmb}$~\cite{cascade,em-cascade},
where $\varepsilon_{\rm cmb} = 6.35 \times 10^{-4}$ eV is the mean energy of
CMB photons.

We can account for the absorption of cascade photons on the EBL,
integrating their production rate $Q_\gamma^{\rm cas}(E)$ over the volume of
the universe,
\begin{equation}
\label{Jabs}
J_{\rm abs}^{\rm cas}(E) =
\frac{c}{4\pi} \int dV \frac{Q_\gamma^{\rm cas}(E)}{4\pi r^2 c}
\exp \left(- \frac{r}{l_{\rm int}(E)} \right) \,.
\end{equation}

Integrating the rate from $r=0$ up to $cH_0^{-1}$ we obtain the relation
between the absorbed flux $J_{\rm abs}^{\rm cas}(E)$ and the unabsorbed flux
$J_\gamma^{\rm cas}(E)$,
\be
\label{JabsJgam}
J_{\rm abs}^{\rm cas}(E) = J_\gamma^{\rm cas}(E) \:
\frac{l_{\rm int}(E)}{c H_0^{-1}} \:\left[1 - \exp \left( 
- \frac{cH_0^{-1}}{l_{\rm int}(E)} \right)\right].
\ee
Here, $H_0$ denotes the present value of the Hubble parameter.

\begin{figure}[!ht]
\begin{center}
\includegraphics[width=0.95\columnwidth]{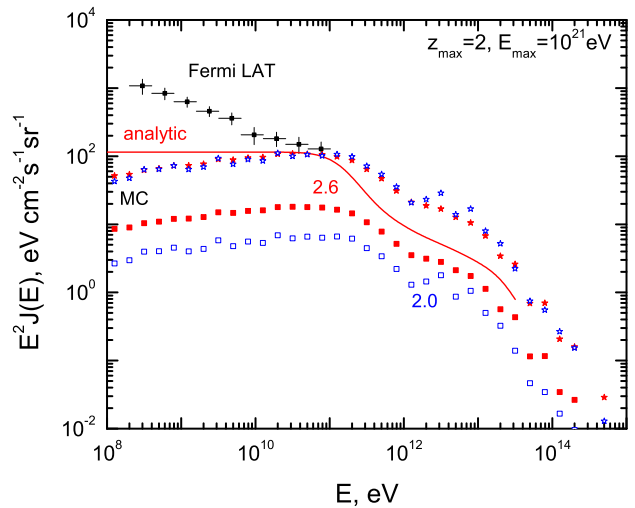}
\caption{Fermi-LAT EGRB spectrum (black circles with with error bars) 
in comparison with maximally allowed fluxes given by analytical 
(solid red line) and MC calculations (red stars for $\alpha_g=2.6$, 
and blue stars $\alpha_g=2.0$). All three curves are normalized by 
the highest energy point of the Fermi spectrum. The MC EGRB fluxes are 
calculated for the following values of the parameters: 
$E_{\max}=10^{21}$\,eV, $z_{\max}=2$, $m=0$  and $\alpha_g=2.6$ or $2.0$. 
Also shown  two MC spectra with the same parameters as above, 
but normalized to the HiRes proton spectrum (red boxes for $\alpha_g=2.6$ 
and blue boxes for  $\alpha_g=2.0$ ). The plot illustrates the 
universality of  the cascade spectrum and reasonably good agreement 
between  MC and analytical results.
}
\label{fig1}
\end{center}
\end{figure}

The cascade energy density $\omega_{\rm cas}$ at the present
epoch is calculated as
\be
\omega_{\rm cas} =   \frac{4 \pi}{c} \int \!dE\; E\,
J_{\rm abs}^{\rm cas}(E) \,.
\ee
In Fig.~\ref{fig1}, we show the measurement of the EGRB by
Fermi-LAT~\cite{fermi} (black circles with error bars)  together
with the maximally allowed photon flux (solid red line) derived
analytically. More precisely, we have determined the maximally allowed 
photon flux requiring that the curve just touches the lower end of 
the error bars of the Fermi-LAT data. 
The corresponding bound on the cascade energy density is 
$\omega_{\rm cas}^{\max}=5.8\times 10^{-7}$\,eV/cm$^3$.

The bound on $\omega_{\rm cas}^{\max}$ derived by us
can be used to select in a simple way viable UHECR models.

We limit our consideration to pure proton-composition models,
which are described by the generation index $\alpha_g$, the maximum 
acceleration energy $E_{\max}$ and the cosmological evolution of the
sources parametrized by  $(1+z)^m$ with fixed $m$ and maximal redshift  
$z_{\max}$. The quantity of interest, the space density of
protons $n_p(E,z)$ at each cosmological epoch, is calculated as
in Ref.~\cite{BGG-dip} in the continuous energy-loss approximation. 
To evaluate the role of fluctuations in $p+\gamma \to \pi + $all 
scattering, the density of protons in Ref.~\cite{BGG-dip} was computed 
also solving the kinetic equation (16). It was found that for 
$E_{\max}=1\times 10^{21}$~eV both methods give practically 
identical results, while for
$E_{\max}=1\times 10^{23}$~eV the difference does not exceed $15\%$.
In the calculations of this work, the UHE proton fluxes at $z=0$ are  
normalized to fit the HiRes spectra~\cite{Abbasi:2007sv}.

The UHE diffuse neutrino flux  at highest energies 
depends mainly on $\alpha_g$ and $E_{\max}$. 
The generation index is limited as $2.0 \lsim \alpha_g \lsim 2.7$, by the 
following reason: Let us choose first the minimum  possible index 
$\alpha_g=2.0$. In this case, the calculated extragalactic UHECR flux is 
very flat and can explain the observed spectrum only above 
$E \sim (0.5 - 1)\times 10^{19}$~eV, i.e.\ above the ankle. 
Increasing $\alpha_g$ decreases the predicted transition energy between
Galactic and extragalactic UHECRs, until for $\alpha_g \approx 2.6 - 2.7$ 
this energy becomes lower than $1\times 10^{18}$~eV, where, 
as observations show, heavy nuclei dominate. 

UHECR models allowed by Fermi-LAT data, i.e.\ leading to a 
cascade energy density $\omega_{\rm cas}$ below 
$\omega_{\rm cas}^{\max}=5.8\times 10^{-7}$\,eV/cm$^3$, are characterized by a
low value of $E_{\max}$ or weak source evolution.
For each given model, we calculate $\omega_{\rm cas}$ as
\be
\omega_{\rm cas} = \int \frac{dt\,dE\;}{1+z}\, E\,\beta_{0,\rm em}[(1+z)E]
\,n_p(E,z) \,,
\label{omega-cas}
\ee
where $n_p$ is the (physical) density of protons at redshift $z$, 
$\beta_0(E)=(1/E)(dE/dt)$ is the relative rate of energy loss of 
a proton with energy $E$ at $z=0$, and $\beta_{0,\rm em}$ denotes the 
relative rate of energy injected by protons into electromagnetic
cascades due to pair production and pion production ($p\gamma \to 
\pi^{\pm} \to e^{\pm}$ and $p\gamma \to \pi^0 \to \gamma$) at $z=0$. 
In Table~1,
we report the numerical values obtained for $\omega_{\rm cas}$ for different
values of the maximal energy of acceleration $E_{\max}$, exponents $\alpha_g$
of the generation spectrum and the maximal
redshift $z_{\max}$. The UHE proton fluxes at $z=0$ are  
normalized to the HiRes data~\cite{Abbasi:2007sv}. In Table~1 we show 
also the ratio of
the contribution from pair production to the one from pion production.
The cases with different $\alpha_g$ in Table~1 correspond to
different transition energies from galactic to extragalactic cosmic
rays, increasing from $E \sim 1\times 10^{18}$~eV for $\alpha_g = 2.6$ to 
$E \sim 5\times 10^{18}$~eV for $\alpha_g = 2.0$. Table~1 gives 
examples of UHECR models that are allowed and that are forbidden by the 
Fermi-LAT data.

The upper part of Table~1 (no-evolution case) presents the allowed
models with $\omega_{\rm cas} < 5.8\times 10^{-7}$\,eV/cm$^3$.
In the two lower parts the cosmological evolution of UHECR sources is 
included, assuming that
the product of the comoving source density $n_s$ and the source luminosity 
$L_s$ evolves as $n_s(z)L_s(z)=n_0L_0(1+z)^m$. 
The middle part contains   both allowed
and marginally allowed evolutionary models.
 The large neutrino fluxes at highest energies, favorable for
 detection by the  JEM-EUSO instrument and by radio methods, 
 are expected in the models with flat spectra ($\alpha_g=2.0$)
  and large $E_{\max}$. Examples of such models  allowed by $\omega_{\rm cas}$
  are also given in  Table~1, most notably the one in the  first row 
  of the middle part. 
  On the other hand, large neutrino fluxes at energies up 
  to $1\times 10^{17}$~eV, which are favorable for IceCube detection, 
  do not require large $E_{\max}$ but strong evolution.
One can find such models in Table~1, too, among the allowed models.
In the lower part of Table~1 three examples for evolutionary models 
forbidden by the Fermi-LAT   data are shown.

Note that apart from the cascade radiation, one should expect various 
additional contributions to the measured Fermi flux and these
contributions lower $\omega_{\rm cas}^{\max}$ further. Among them are 
photons from unresolved~\cite{stecker} or dead~\cite{neronov} active 
galactic nuclei (AGN)  and from other galaxies. 
Dark matter annihilations and/or decays in the extended DM Galactic
halo or beyond can give another contribution to the Fermi flux. 
Subtracting these processes would strengthen the upper limit on the 
neutrino flux derived below, and thus our result is conservative.
On the other hand, extragalactic magnetic fields with strength above 
1~nG play the opposite role:  High-energy cascade electrons loose energy 
radiating predominantly synchrotron photons, which are not
detected by Fermi-LAT and thus the energies of these electrons do not
contribute to $\omega_{\rm cas}$. We show below that in the case 
of magnetic field strengths less than 1~nG  this correction on the limit 
for UHE neutrino flux is small.

\begin{center}
\begin{table}
\caption{
The energy density $\omega_{\rm cas}$ of the cascade radiation
produced by UHE protons normalized to HiRes data.}
\smallskip
\begin{tabular}{ ||c|c|c|c|c|c||}
\hline
\multicolumn{6}{||c||}{No evolution, allowed models} \\ \hline
$\;m\;$ &$E_{\max}$ & $\alpha_g$ & $z_{\max}$ & $\omega_{\rm cas}$ [eV/cm$^3$] &
$\omega_{\rm cas}^{e^+ e^-}/\omega_{\rm cas}^{\rm \pi}$ \\ \hline
$0$ &$10^{21}$ & $2.0$ & $2$ & $4.0\cdot 10^{-8}$ & $2.41$ \\ \hline
$0$ &$10^{21}$ & $2.7$ & $2$ & $1.2\cdot 10^{-7}$ & $24.5$ \\ \hline
$0$ &$10^{22}$ & $2.0$ & $2$ & $5.1\cdot 10^{-8}$ & $1.2$ \\ \hline
$0$ &$10^{22}$ & $2.7$ & $2$ & $1.2\cdot 10^{-7}$ & $20.8$ \\ \hline
$0$ &$10^{22}$ & $2.0$ & $3$ & $5.5\cdot 10^{-8}$ & $1.2$ \\ \hline
$0$ &$10^{22}$ & $2.7$ & $3$ & $1.4\cdot 10^{-7}$ & $22.0$ \\ \hline
\multicolumn{6}{||c||}{With evolution, allowed models} \\ \hline
$2.5$ & $10^{22}$ & $2.0$ & $4$ & $2.7 \cdot 10^{-7}$ & 1.46 \\ \hline
$3$ & $10^{21}$ & $2.5$ & $2$ & $4.3 \cdot 10^{-7}$ & 14.7 \\ \hline
$3$ & $10^{22}$ &$2.4$ & $3$ & $5.8 \cdot 10^{-7}$ & 8.4 \\ \hline
\multicolumn{6}{||c||}{Models excluded by $\omega_{cas}$} \\ \hline
$3.5$ & $10^{23}$ & $2.3$ & $3$ & $7.4 \cdot 10^{-7}$ & 4.9 \\ \hline
$4$ & $10^{21}$ &$2.5$ & $2$ & $8.6 \cdot 10^{-7}$ & 15.3 \\ \hline
$4$ & $10^{22}$ & $2.0$ & $3$ & $7.6\cdot 10^{-7}$ & 1.5 \\ \hline
\end{tabular}
\end{table}
\end{center}

\section{Monte Carlo simulation}%
%
In addition to the analytical treatment, we obtain the EGRB spectrum
based on a Monte Carlo simulation of the cascade development.
We generate CR sources from a homogeneous source distribution
up to a maximal redshift $z_{\max}$.
Assuming the proton injection spectrum in the form
$dN/dE\propto E^{-\alpha_g}\theta(E-E_{\max})$, we propagate the UHE protons accelerated in the
sources through the extragalactic space, using the Monte Carlo code
described in \cite{Kachelriess:2004pc}, until their energy
is below the threshold for $e^+e^-$ pair production, $E_{\min}\ap 10^{18}\,$eV,
or until they reach the Earth. For the simulation of pion production we use
SOPHIA~\cite{sophia}, while   $e^+e^-$ pairs are injected according to
the continuous energy losses and their mean energy calculated in
Ref.~\cite{BGG-dip,Kelner:2008ke}

We follow the evolution of electromagnetic cascades using the Monte Carlo
code introduced in Ref.~\cite{Kachelriess:2008qx} and the best-fit
model of \cite{kneiske04} for the EBL energy
density. The MC procedure provides an one-dimensional description of the
cascade development, taking into account the pair production and IC
processes as well as adiabatic energy losses.
Extragalactic magnetic fields with average strengths close to the
upper limit $B\sim 1$\,nG
have a {\rm small} influence (of order 20\%) on the resulting EGRB. 
We will discuss further this result below.

\begin{figure}
\begin{center}
\includegraphics[width=0.95\columnwidth]{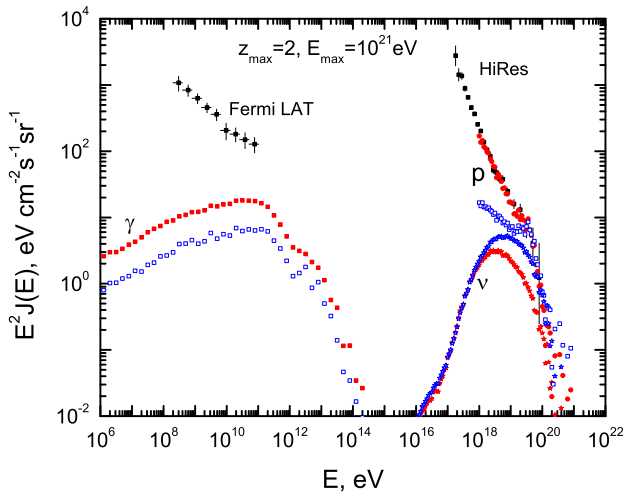}
\caption{Fermi-LAT data (black circles) for the EGRB and UHECR data from
HiRes (dots) together with  UHE neutrino (stars) and photon (boxes)
fluxes for  $E_{\max}=10^{21}$\,eV, $z_{\max}=2$, $m=0$ 
and
$\alpha_g=2.0$ (blue, open) and
$\alpha_g=2.6$ (red, filled symbols).}
\label{fig2}
\end{center}
\end{figure}

In Fig.~\ref{fig1}, the cascade fluxes are shown for two UHECR models. The
curve marked as $\alpha_g=2.6$ (red boxes) gives the
cascade flux for the non-evolutionary ($m=0$) dip model \cite{BGG-dip}
with  $E_{\max} = 1\times 10^{21}$~eV and $z_{\max}=2$
normalized to HiRes data. The other curve marked as $\alpha_g=2.0$
is shown for the ankle model with a transition from galactic to
extragalactic cosmic rays at $5\times 10^{18}$~eV for the same values
of $E_{\max}$ and $z_{\max}$. From Fig.~\ref{fig1}, one can see 
that both models are allowed by the cascade limit.

The MC simulation allows us to test the {\em universality\/} of the 
cascade spectrum.
If a cascade is initiated by a photon or an electron of very high energy, the
energy spectrum of the resulting cascade photons depends only weakly on the 
energy of the primary particle for a sufficiently large number of cascade
steps. This
universality is obviously broken for the primaries injected close enough to an
observer, if the  distance is of the order of the absorption length (see   
Eq.~\ref{Jabs}). In Fig.~\ref{fig1} we plot the MC cascade spectra 
with $\alpha_g = 2.6$ and $\alpha_g = 2.0$ normalizing them by the  
highest energy point of the Fermi spectrum (red and blue stars in 
Fig.~\ref{fig1}). The comparison of the three theoretical spectra  at
energies below the minimal absorption energy $\eps_a$ shows that the  
cascade spectrum is indeed quite universal. The shape of the cascade photon 
spectra from 
the Monte Carlo simulation agrees reasonably well with the one 
analytically calculated, with a somewhat harder photon flux obtained 
with the Monte Carlo method in the plateau region, $J(E)\propto E^{-1.95}$.  
As a result, the maximal cascade energy density $\omega_{\rm cas}^{\max}$ 
obtained using the Monte Carlo simulation
is 30\% smaller than in the analytic calculations.

In Fig.~\ref{fig2}, we show the obtained UHECR, neutrino and photon
fluxes together with data from HiRes and Fermi-LAT for the two
cases $\alpha_g=2.0$ (blue) and 2.6 (red). We use again $E_{\max}=10^{21}$\,eV,
$z_{\max}=2$ and normalize the UHECR results to the  HiRes observations.
While the dip model fits the HiRes data with $\chi^2=19.5$ for
$d.o.f.=19$,  the ankle model cannot explain the HiRes data below 
$1\times 10^{19}$\,eV without an additional component.
Clearly, the dip scenario without evolution and with modest values
of $E_{\max}$ and  $z_{\max}$ is well compatible with the Fermi data
(see Fig.~\ref{fig1}). 
The ankle scenario with $\alpha_g=2.0$ has a lower flux
of cascade gamma-radiation and is viable too.

\begin{figure}[t]
\begin{center}
\includegraphics[width=0.9\columnwidth]{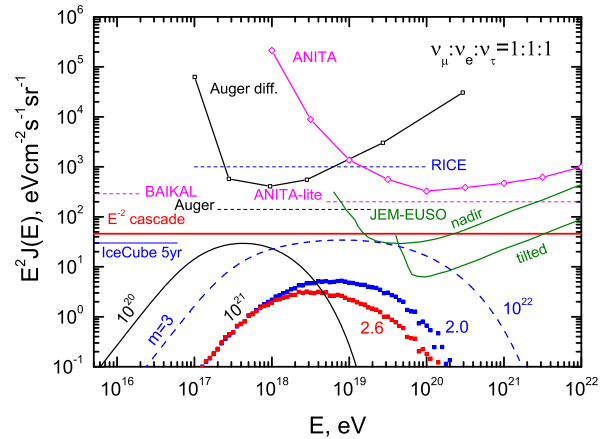}
\caption{Upper limits on the all-flavor UHE neutrino flux and expected
sensitivities \cite{upper-limits} together with the cascade limit 
(``$E^{-2}$ cascade'').
Also shown are realistic fluxes of cosmogenic neutrinos marked
by their spectral index $\alpha_g=2.6$ (dip model) and
$\alpha_g=2.0$ (ankle model) together with neutrino fluxes
optimized for detection by IceCube and JEM-EUSO 
(as described in Section IV), which marked in the Figure by their 
respective  $E_{\max}$ values in eV ($10^{20}$ and $10^{22}$).      
}
\label{fig3}
\end{center}
\end{figure}

\begin{figure}[t]
\begin{center}
\includegraphics[width=0.9\columnwidth]{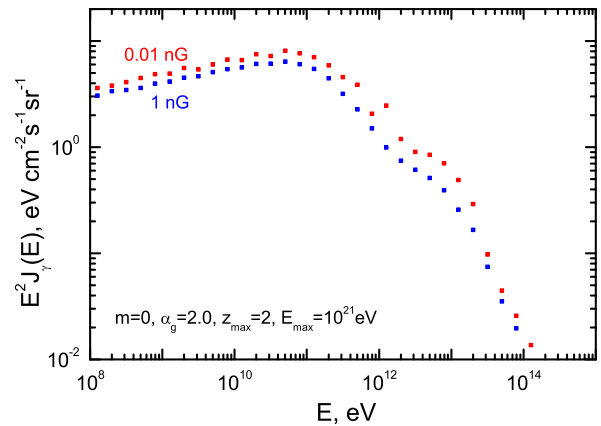}
\caption{Photon fluxes from the Monte Carlo simulation
for different magnetic field strengths $B= 0.01$ and 1\,nG 
with  $E_{\max}=10^{21}$\,eV, $z_{\max}=2$, $m=0$ and $\alpha_g=2.0$.}
\label{figB}
\end{center}
\end{figure}

\section{The cascade bound on UHE neutrinos}%
%
This is the most general bound on the UHE neutrino flux,
based only on the production of electromagnetic cascades, which inevitably
accompany the production of pions responsible for the neutrino flux
\cite{cascade,em-cascade}. 
It is based on the approximate equality of the total energy release 
to neutrino radiation (through $p\gamma \to \pi^{\pm} \to \nu$
and to the cascade radiation (through 
$p\gamma \to \pi^{\pm} \to e^{\pm}$ and $p\gamma \to \pi^0 \to \gamma$)
in pion production process.  
The  upper limit on the integral flux  $J_{\nu}(>E)$  of neutrinos
of all flavors is given by the following chain of inequalities,
$$
\omega_{\rm cas}^{\max} > \omega_{\rm cas}^{\pi}>
\frac{4\pi}{c}\int_E^{\infty} \!\!\! E'J_{\nu}(E')dE'>
\frac{4\pi}{c}EJ_{\nu}(>E),
$$
where $\omega_{\rm cas}^{\max}$ and $\omega_{\rm cas}^{\pi}$ are  the
energy density of the cascade radiation allowed by the Fermi data
and that produced only by pions, respectively. For the sake of comparison
with experimental  upper bounds, where a $E^{-2}$ neutrino spectrum
is usually assumed, we give  the upper
limit for the differential cosmogenic neutrino flux
of three neutrino flavors
with a $E^{-2}$ spectrum
and as function of the ratio of energy densities of pair- and pion-produced
cascades $\omega_{\rm cas}^{e^+e^-}/\omega_{\rm cas}^{\pi}$,
\be
E^2J_\nu (E) \leq \frac{c}{4\pi}\; \frac{\omega_{\rm cas}^{\max}}
{\ln (E_{\rm max}/E_{\rm min})}\;
\frac{1}{1+\omega_{\rm cas}^{e^+e^-}/\omega_{\rm cas}^{\pi}} \,.
\label{cosmogenic-bound}
\ee
This limit is plotted in Fig.~\ref{fig3} as a red line labeled
'$E^{-2}$~cascade' together with existing upper limits from various
experiments and the expected sensitivity of
IceCube and JEM-EUSO~\cite{upper-limits,Ice}.
Equation~(\ref{cosmogenic-bound}) 
gives the general upper limit on the neutrino flux using the $E^{-2}$ 
assumption. However, each particular model for cosmogenic neutrinos
can be checked for consistency with the Fermi bound
straightforwardly, as described in Section II. Namely, $\omega_{\rm cas}$
can be calculated from Eq.~(\ref{omega-cas}) and compared with 
$\omega_{\rm cas}^{\max}=5.8\times 10^{-7}$~eV/cm$^3$. 

We discuss now the impact of magnetic fields on the cascade limit. In the 
presence of magnetic fields, the energy of electrons is partly dissipated 
in the form of synchrotron radiation. 
The critical energy $E_e^{\rm cr}$ of electrons above which 
synchrotron energy losses dominate is determined by the relation 
$(dE/dt)_{\rm syn}=(dE/dt)_{\rm IC}$,
where the indices 'syn' and 'IC' are related to synchrotron and 
IC losses, respectively. 

In the case of a single electron with energy $E_0 > E_e^{\rm cr}$, 
the usual electromagnetic cascade is suppressed until the electron energy 
drops below
$E_e^{\rm cr}$. In this case, the total cascade energy is reduced by the
factor $E_e^{\rm cr}/E_0$ compared to the case without magnetic
field.  However, the cascade is initiated  by many
electrons with production spectrum $\propto E_e^{-2}$, as required for 
a $E_\nu^{-2}$-upper limit. Then the cascade energy density 
$\omega_{\rm cas}(E_e)dE_e$ is proportional to $E_\nu J_\nu (E_\nu)dE_\nu$,
and the ratio of the cascade energy density in presence of a magnetic field 
$\omega_{\rm cas}^B$ and in its absence $\omega_{\rm cas}$ is given by 
\begin{equation}
\frac{\omega_{\rm cas}^B}{\omega_{\rm cas}}=
\frac{\ln (E_e^{\rm cr}/E_e^{\min})}{\ln (E_e^{\max}/E_e^{\min})} \,.
\label{ratio-omega1}
\end{equation}
For the case of a strong magnetic field $B \sim 1$~nG one obtains 
$E_e^{\rm cr} \sim 2\times 10^{18}$~eV. Taking 
$E_{\max} \sim 1\times 10^{21}$~eV and $E_{\min} \sim 1\times 10^9$~eV, 
we find $\omega_{\rm cas}^B/\omega_{\rm cas} =0.78$, i.e.\ only 
$22\%$ of the cascade energy is lost due to synchrotron radiation. 

The case considered above corresponds to the $E^{-2}$ upper limit shown in 
Fig.~\ref{fig3}. For cosmogenic neutrinos produced by protons interacting with
CMB the fraction of energy lost from the cascade is less because of
the steeper generation spectrum $Q_p(E) \propto E^{-\beta_g}$ with 
$\beta_g \sim 2.3 - 2.7$. As a result the cascade energy is produced
mainly by low-energy electrons, for which IC dominates. The 
ratio (\ref{ratio-omega1}) is given now by  
\begin{equation}
\frac{\omega_{\rm cas}^B}{\omega_{\rm cas}}=
\frac{1- (E_e^{\rm cr}/E_{\min})^{-(\beta_g -2)}} 
{1- (E_{\max}/E_{\min})^{-(\beta_g -2)}} ,
\label{ratio-omega2}
\end{equation}
and is for all practical cases very close to 1. 
Our photon fluxes from MC simulations with $B=0.01$ and 1\,nG 
displayed in Fig.~\ref{figB} show 
indeed only minor differences, compatible with the analytical estimate
given above.

\section{Cosmogenic UHE neutrino fluxes}%

We discuss now the cosmogenic neutrino fluxes compatible
with the two conditions that the parent proton fluxes provide a good 
fit to the HiRes data and that the resulting EGRB respects the cascade 
bound. The latter is imposed by requiring that $\omega_{\rm cas}$ calculated 
with the help of Eq.~(\ref{omega-cas}) is smaller than $\omega_{\rm cas}^{\max}$ 
deduced from the Fermi data.   

We begin with our reference models, given by the dip ($\alpha_g=2.6$) 
and ankle ($\alpha_g=2.0$)  model 
normalized to the HiRes data, and using $E_{\max}=1.0\times 10^{21}$~eV,
$z_{\max}=2$ and non-evolution. These are  conservative models which give 
in case of $\alpha_g=2.6 - 2.7$ the lowest neutrino fluxes for the 
proton-dominated mass composition. These fluxes labeled in 
Fig.~\ref{fig3} by the generation indices 2.0 and 2.6 are shown in 
the central part of the figure. They are undetectable by Auger and 
the planned detector JEM-EUSO.  

\begin{figure*}[!t]
\begin{center}
\includegraphics[width=0.98\columnwidth]{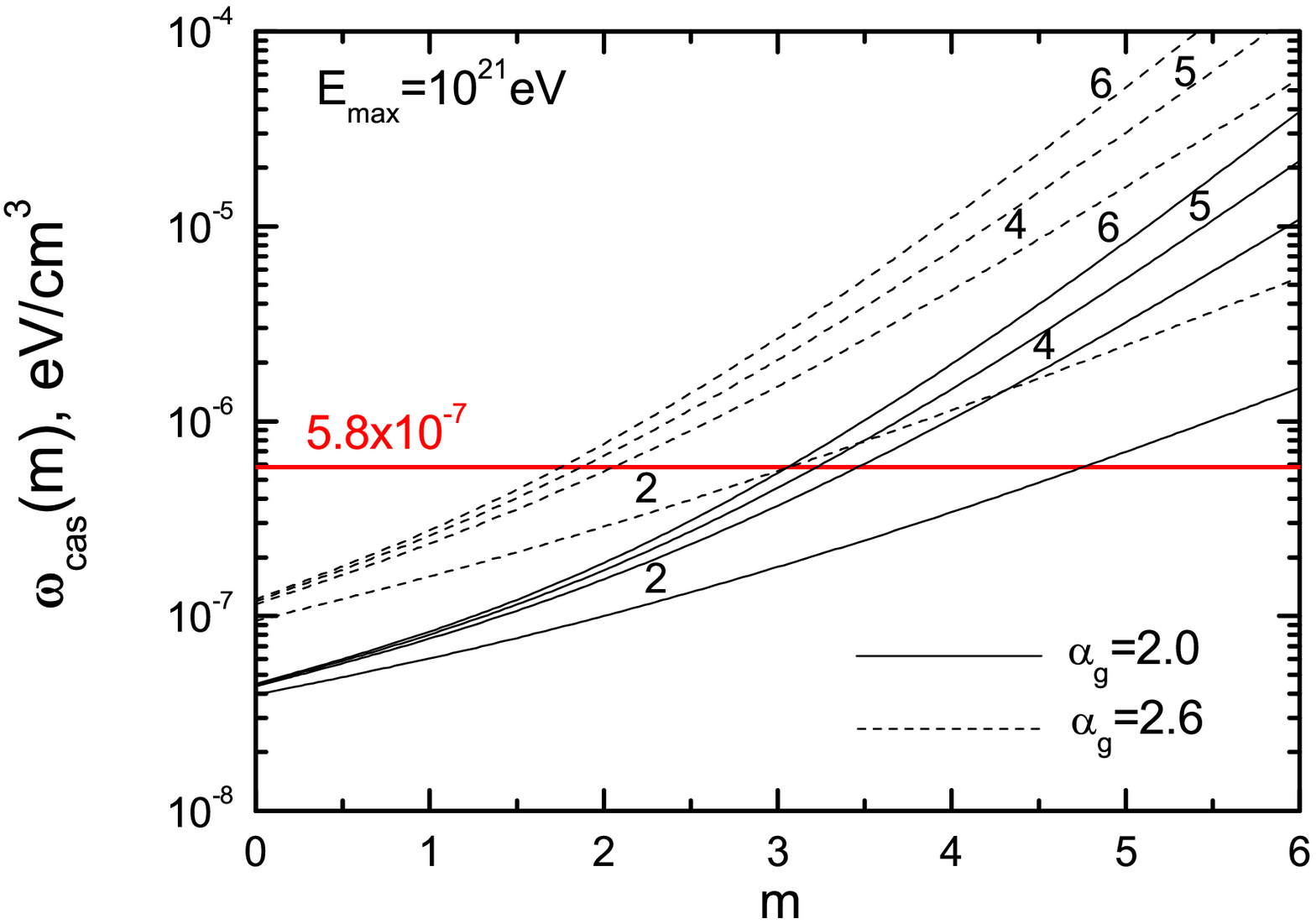}
\includegraphics[width=0.93\columnwidth]{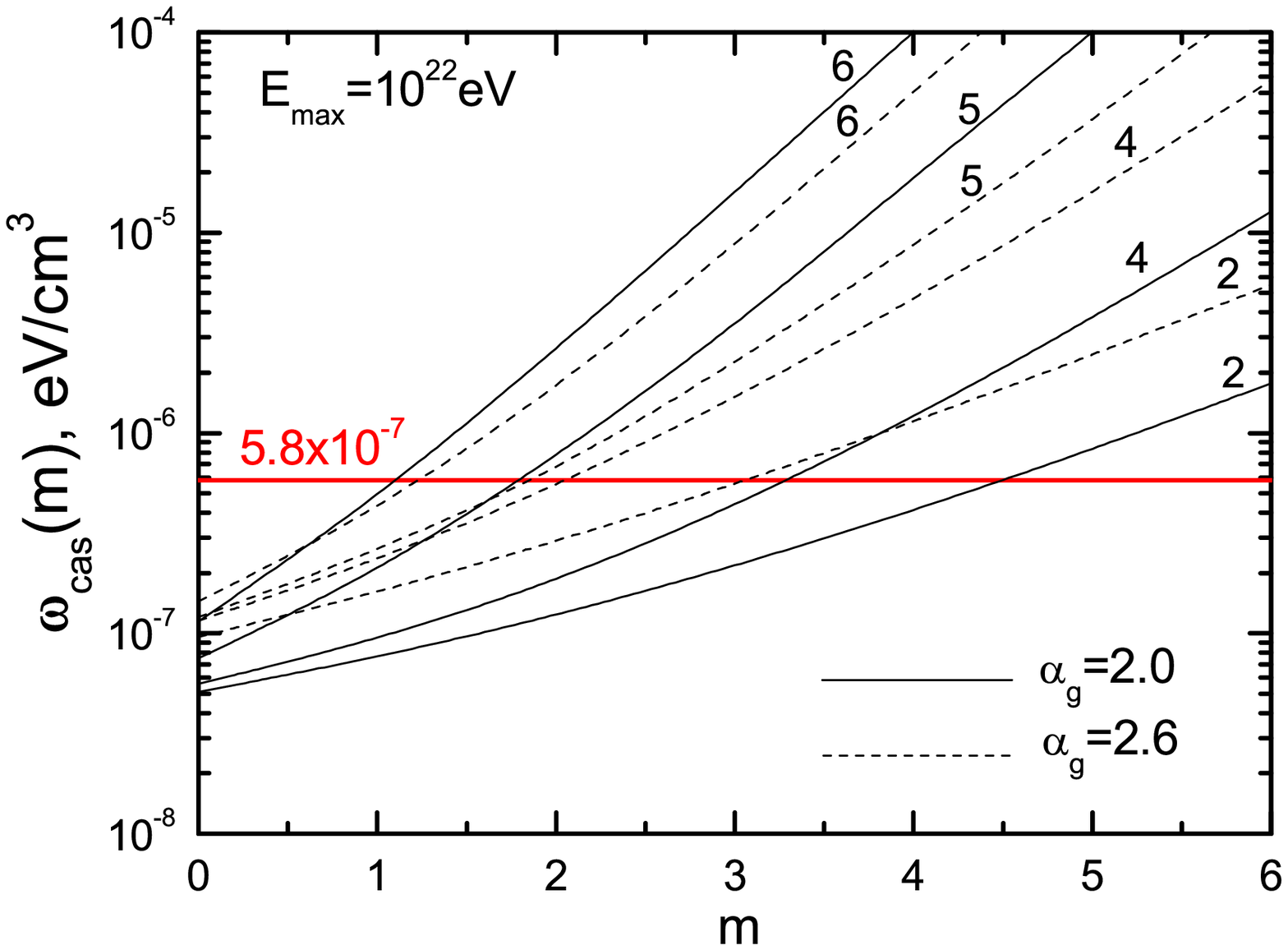}
\caption{Range of allowed evolution parameters, $m$ and $z_{\max}$,
for extended reference models with fixed 
$E_{\max}=1\times 10^{21}$~eV (left panel) and 
$E_{\max}=1\times 10^{22}$~eV (right panel). The cascade energy
density $\omega_{\rm cas}$ is shown as function of $m$ by the 
solid lines for the ankle model ($\alpha_g=2.0$), and dashed lines 
for the dip model ($\alpha_g=2.6$). The numbers on the lines 
show $z_{\max}$. The allowed parameters correspond to part of the
curves below $\omega_{\rm cas}^{\max}=5.8\times 10^{-7}$~eV/cm$^3$
shown by the red horizontal line.
}
\label{fig5}
\end{center}
\end{figure*}

Next we extend our reference models, allowing larger  
$E_{\max}$ and cosmological evolution. This results 
in higher neutrino fluxes, limited however still by  
$\omega_{\rm cas}$. The bounds on the parameters of these
models are shown in Fig.~\ref{fig5}. The two panels of this figure
show for $E_{\max}=1\times 10^{21}$~eV (left panel) and 
$E_{\max}=1\times 10^{22}$~eV (right panel), how the two parameters describing 
source evolution, $m$ and $z_{\max}$, are limited 
by $\omega_{\rm}^{\max}=5.8\times 10^{-7}$~eV/cm$^3$.
Generally, strong evolution $(m, z_{\max})$ with  
$m \gsim 3$ and $z_{\max} \gsim 4$ is
excluded, in accordance with the cases presented in Table~1. 
The evolution in the dip model ($\alpha_g=2.6$) is restricted stronger, 
because of the increased contribution from  $e^+e^-$ pair-production.

The cosmogenic neutrino flux can become detectable
only in the case of source evolution and large $E_{\max}$.
Two extreme models of such neutrino fluxes  are shown in the lower-right
and lower-left corners  of Fig.~\ref{fig3}.
Both models respect the bound from the observed UHECR flux and
from $\omega_{\rm cas}^{\max}$ derived from the Fermi-LAT data,
but use extreme values for the model parameters.
Choosing the parameters for the model in the
lower-right  corner  (the curve marked $10^{22}$)
we try to reach the sensitivity of JEM-EUSO. Since
a soft spectrum increases $\omega_{\rm cas}$, we choose 
{\rm the hard spectrum with} $\alpha_g=2.0$,
while $E_{\max}$ should be as large as possible. By other words we 
search for the extension of the ankle reference model with 
allowed evolution and large $E_{\max}$.  We choose
$E_{\max}=1\times 10^{22}$~eV, with  $z_{\max}=2$ and
evolution parameter $m=3$. Normalized to the HiRes data, this model has
$\omega_{\rm cas}= 3.3 \times 10^{-7}$~eV/cm$^3$, i.e.\ is somewhat below the
cascade limit (see also Fig.~\ref{fig5}). For such values, the neutrino flux is
marginally detectable by JEM-EUSO.

In the lower-left corner   (the curve marked $10^{20}$) we  
aim to cosmogenic neutrino detection by
IceCube. Here we should increase the low-energy tail
of the neutrino flux and suppress the pair-produced  cascade
radiation. To that end, we use $\alpha_g=2.0$ with strong
evolution to enhance the flux of low-energy neutrinos. The maximum
acceleration energy can be low, e.g.\ $E_{\max} =1\times 10^{20}$~eV.
Moreover, we choose evolution with $m=3.0$ and $z_{\max}=6.0$, which
results in $\omega_{\rm cas}=5.5\times 10^{-7}\,{\rm eV/cm}^3
\approx \omega_{\rm cas}^{\max}$. As our calculations show,  the flux 
is only marginally detectable by IceCube even for these extreme parameters.

The two models above demonstrate that even for extreme assumptions
cosmogenic neutrinos remain undetectable by existing  detectors 
such as Auger, and could be only marginally observed
by IceCube and by future detectors JEM-EUSO and Auger-North
(with sensitivity to neutrinos 5--6 times higher than Auger-South).

The observation of radio emission from neutrino-induced air showers
provides an effective method for the detection of low fluxes of cosmogenic 
neutrinos from the highest energy part of their spectrum.
The upper limit on UHE cosmogenic neutrino flux from the most
restrictive experiment of this type, ANITA, is shown in Fig.~3
(Gorham {\it et al.} \cite{upper-limits}). 
Recently, several particles with energies above $1\times 10^{19}$~eV
have been detected there \cite{ANITA}. The high energy threshold is a
disadvantage of this method. In the recently proposed ARIANNA detector
\cite{ARIANNA}, the threshold might be lowered to about $10^{17}$~eV 
while monitoring 900~km$^2$ of Antarctic ice.

A very sensitive instrument for UHE neutrino detection has been proposed
 in the project LORD (Lunar Orbital Radio Detector) \cite{lord}, 
where a detector on a lunar
satellite can observe the neutrino-produced radio-signal from lunar
regolith. The sensitivity of this instrument, as estimated by the
authors of the project, should be sufficient for the measurement of
the  cosmogenic neutrino fluxes shown in Fig.~\ref{fig3} by curves 
$10^{21}$. 

Before concluding, we would like to compare the results of this investigation 
to the ones of Ahlers {\it et al.}~\cite{ahlers} that appeared 
after ours in the arXiv. While the main goal of our work was to derive
an upper limit on the cosmogenic neutrino flux, the authors of 
Ref.~\cite{ahlers} aimed at exploring the allowed parameter space of
UHECR models, notably of those predicting maximal neutrino fluxes. 
These authors used as their criterion for the rejection of UHECR models 
$\omega_{\rm cas}^{\max}=5.8\times 10^{-7}$~eV/cm$^3$ from our calculations, 
and thus the derived maximally allowed cosmogenic neutrino fluxes should
coincide. The largest cosmogenic neutrino fluxes presented 
in Fig.~4 of Ref.~\cite{ahlers} are very similar to our fluxes obtained in the
extreme models with strong cosmological evolution (e.g.~the   curve $10^{22}$ 
in Fig.~\ref{fig3}),  both  exceeding our reference cases ($\alpha_g=2.6$ and
 $\alpha_g=2.0$ without evolution) by an order of magnitude
 at $E \sim   10^{18}\div 10^{19}$~eV. It is noteworthy that a much stronger
 cosmological evolution was considered in the calculations of 
 Ref.~\cite{ahlers}. Among   other differences, the authors of
Ref.~\cite{ahlers} assumed that the IceCube sensitivity extends up to
$10^{19}$\,eV, while we used $E_{\max}=10^{17}$\,eV following 
Ref.~\cite{Ice}.

\section{Summary}%
%
We have used a recent measurement of the EGRB by Fermi-LAT
to constrain models for UHECR and cosmogenic UHE neutrinos
and to demonstrate that the latter are  not detectable
with the present experimental sensitivity. Both the dip and ankle
model without or with weak evolution are consistent with the
Fermi-LAT measurement of the EGRB. The cosmogenic neutrino flux is strongly
limited by the new upper cascade bound and  undetectable 
for a conservative choice of parameters by Auger-North and JEM-EUSO. 
Only for an extreme
set of parameters, $E_{\max} \gsim 1\times 10^{22}$~eV and
$\omega_{\rm cas} \sim \omega_{\rm cas}^{\max}$,  the cosmogenic flux
is marginally detectable by JEM-EUSO. To achieve the observation of 
cosmogenic
neutrinos for less extreme parameters, the  detection
threshold of JEM-EUSO (in the tilted mode) must be lowered down to
$1\times 10^{19}$~eV and the sensitivity of Auger-North should be
increased by factor $\sim 20$ in comparison with Auger-South.
The further development of radio-detection methods gives another hope 
for detection of small fluxes of cosmogenic neutrinos.

The results of our paper emphasize the necessity to develop 
more sensitive methods for the detection of cosmogenic neutrinos.

\section*{Acknowledgments}%
%
VB and AG are grateful to the ICTP, Trieste for hospitality.
S.O.\  acknowledges a Marie Curie IEF fellowship.
This work was partially supported by the program Romforskning of
Norsk Forskningsradet.



\begin{thebibliography}{00}

\bibitem{GZK}
K.~Greisen,
Phys.\ Rev.\ Lett.\  {\bf 16}, 748 (1966);
G.~T.~Zatsepin and V.~A.~Kuzmin,
JETP Lett.\  {\bf 4}, 78 (1966)
[Pisma Zh.\ Eksp.\ Teor.\ Fiz.\  {\bf 4}, 114 (1966)].


\bibitem[Abbasi(2007)]{Abbasi:2007sv}
 R.~Abbasi {\it et al.}  [HiRes Collaboration],
 Phys.\ Rev.\ Lett.\  {\bf 100}, 101101 (2008).


\bibitem{Abraham:2008ru}
 J.~Abraham {\it et al.}  [Pierre Auger Collaboration],
 Phys.\ Rev.\ Lett.\  {\bf 101}, 061101 (2008).


\bibitem{BZ}
V.~S.~Beresinsky and G.~T.~Zatsepin,
Phys.\ Lett.\ B {\bf 28}, 423 (1969);
Sov.\ J.\ Nucl.\ Phys.\ {\bf 11}, 111 (1970).

\bibitem{BG88}
V.~S.~Berezinsky and S.~I.~Grigor'eva,
Astron.\ Astrophys.\  {\bf 199}, 1 (1988).

\bibitem{BGG-dip}
V.~Berezinsky, A.~Z.~Gazizov and S.~I.~Grigorieva,
Phys.\ Rev.\  D {\bf 74}, 043005 (2006);
astro-ph/0210095.


\bibitem{cascade}
V. S. Berezinsky and A. Yu. Smirnov,
Astrophys.\ Sp.\ Sci.\ {\bf 32} 461 (1975);
for recent works see
D.~V.~Semikoz and G.~Sigl,
 JCAP {\bf 0404}, 003 (2004);
Z.~Fodor, S.~D.~Katz, A.~Ringwald and H.~Tu,
 JCAP {\bf 0311}, 015 (2003),
K.~Kotera, D.~Allard and A.~V.~Olinto 
arXiv:1009.1382.

\bibitem{nuclei}
M.~Ave, N.~Busca, A.~V.~Olinto, A.~A.~Watson, and T.~Yamamoto.
Astropart.\ Phys.\ {\bf 23}, 19 (2005),
D.~Hooper, A.~Tailor, and S.~Sarkar,
Astropart.\ Phys.\ {\bf 23}, 11 (2005).


\bibitem{fermi}
A.~A.~Abdo {\it et al.} [Fermi-LAT collaboration],
arXiv:1002.3603 [astro-ph.HE].

\bibitem{em-cascade}
V.~S.~Berezinsky {\it et al.},
Astrophysics of Cosmic Rays (Elsevier, Amsterdam (1990));
V.~S.~Berezinsky, Nucl.\ Phys.\ {\bf B 380}, 478 (1992).

\bibitem{stecker}
F.~W.~Stecker and M.~H.~Salamon, Astroph.~J.~ {\bf 464}, 600 (1996)


\bibitem{neronov}
  A.~Neronov {\it et al.}, 
 Astrophys.\ J.\ Lett.\ {\bf 719}, 130 (2010)


\bibitem{Kachelriess:2004pc}
 M.~Kachelrie\ss\ and D.~Semikoz,
 Astropart.\ Phys.\  {\bf 23}, 486 (2005).

\bibitem{sophia}
A.~M\"ucke {\it et al.},
Comput.\ Phys.\ Commun.\  {\bf 124}, 290 (2000).


\bibitem{Kelner:2008ke}
For a calculation of the energy spectra of electrons see
 S.~R.~Kelner and F.~A.~Aharonian,
 Phys.\ Rev.\  D {\bf 78}, 034013 (2008).


\bibitem{Kachelriess:2008qx}
 M.~Kachelrie\ss, S.~Ostapchenko and R.~Tom\`as,
 New J.\ Phys.\  {\bf 11}, 065017 (2009).


\bibitem[Kneiske et al.(2004)]{kneiske04}
 T.~M.~Kneiske  {\it et al.},
 Astron.\ Astrophys.\  {\bf  413}, 807 (2004).

\bibitem{upper-limits}
 J.~Abraham {\it et al.}  [Pierre Auger Collaboration],
 Phys.\ Rev.\ Lett.\  {\bf 100}, 211101 (2008);
P.~W.~Gorham {\it et al.}  [ANITA collaboration],
{\it et al.}  [ANITA Collaboration],
 arXiv:1003.2961 [astro-ph.HE].
 I.~Kravchenko {\it et al.},
 Phys.\ Rev.\  D {\bf 73}, 082002 (2006);
 N.~Inoue, K.~Miyazawa and Y.~Kawasaki  [JEM-EUSO Collaboration],
 Nucl.\ Phys.\ Proc.\ Suppl.\  {\bf 196}, 135 (2009).

\bibitem{Ice}
 J.~Ahrens {\it et al.}  [IceCube Collaboration],
 Astropart.\ Phys.\  {\bf 20}, 507 (2004);

\bibitem{ANITA}
S.~Hoover {\it et al.} arXiv:1005.0035 [astro-ph.HE].

\bibitem{ARIANNA}
L.~Gerhardt {\it et al.} arXiv:1005.5193 [astro-ph.HE]. 

\bibitem{lord} 
G.~A.~Gusev {\it et al.} Nucl. Instr. Meth. {\bf A 604}, S124 (2009),\\
V.~A.~Ryabov {\it et al.} Nucl. Phys. B (Proc. Suppl)  {\bf 196}, 458
(2009).

\bibitem{ahlers}
M.~Ahlers {\it et al.} Astrop. Phys. {\bf 34}, 106 (2009), arXiv:1005.2620.


\end{thebibliography}
\end{document}